\documentstyle[sprocl]{article}
\def\be{\begin{equation}}
\def\ee{\end{equation}}
\def\bea{\begin{eqnarray}}
\def\eea{\end{eqnarray}}

\begin{document}
\title{MODELS OF NEUTRINO MASS AND INTERACTIONS\\ 
FOR NEUTRINO OSCILLATIONS}
\author{ERNEST MA}
\address{Department of Physics, University of California\\
Riverside, CA 92521, USA}
\maketitle\abstracts{I review the various mechanisms for obtaining nonzero 
neutrino masses.  I focus on the possibility of adding one light singlet 
neutrino to the three known doublet neutrinos $\nu_e$, $\nu_\mu$, and 
$\nu_\tau$ to accommodate the totality of neutrino-oscillation experiments.  
I review the various theoretical suggestions for this singlet.  I discuss 
also a scenario in which there are only the three doublet neutrinos but 
a large anomalous interaction is assumed for $\nu_\tau$ as an attempt to 
explain all present data.}

\section{Neutrino Masses}

In the minimal standard model, under the gauge group $SU(3)_C \times SU(2)_L 
\times U(1)_Y$, the leptons transform as:
\begin{equation}
\left[ \begin{array} {c} \nu_e \\ e \end{array} \right]_L, 
\left[ \begin{array} {c} \nu_\mu \\ \mu \end{array} \right]_L, 
\left[ \begin{array} {c} \nu_\tau \\ \tau \end{array} \right]_L 
\sim (1, 2, -1/2); ~~~ e_R, ~ \mu_R, ~ \tau_R \sim (1, 1, -1).
\end{equation}
There is also the Higgs scalar doublet $(\phi^+, \phi^0) \sim (1, 2, 1/2)$ 
whose nonzero vacuum expectation value $\langle \phi^0 \rangle = v$ breaks 
$SU(2)_L \times U(1)_Y$ to $U(1)_Q$.  Whereas charged leptons acquire masses 
proportional to $v$, the absence of $\nu_R$ implies that $m_{\nu_i} = 0$.  If 
nonzero neutrino masses are desired (which are of course necessary for 
neutrino oscillations), then we must ask ``What is the nature of this mass?" 
and ``What new physics goes with it?"

If $\nu_R$ does not exist, one way to have $m_\nu \neq 0$ is to add 
a Higgs triplet $(\xi^{++}, \xi^+, \xi^0)$.  Each $\nu_L$ then gets a 
Majorana mass.  However, $\langle \xi^0 \rangle$ must be very small, and if 
the lepton number being carried by $\xi$ is spontaneously violated~\cite{1}, 
the decay of $Z$ to the associated massless Goldstone boson (the 
triplet Majoron) and its partner would count as two extra neutrinos.  Since 
the effective number of light neutrinos in $Z$ decay is now measured~\cite{2} 
to be $2.989 \pm 0.012$, the triplet Majoron model is clearly ruled out.

If one $\nu_R \sim (1, 1, 0)$ exists for each $\nu_L$, the most general 
$2 \times 2$ neutrino mass matrix linking $(\bar \nu_L, \bar \nu_R^c)$ to 
$(\nu_L^c, \nu_R)$ is given by
\begin{equation}
{\cal M} = \left[ \begin{array} {c@{\quad}c} m_L & m_D \\ m_D & m_R \end{array} 
\right].
\end{equation}
If $m_L = 0$ and $m_D << m_R$, we get the famous seesaw mechanism~\cite{3}
\begin{equation}
m_\nu \sim {m_D^2 \over m_R}.
\end{equation}
Here, $\nu_L - \nu_R^c$ mixing is $m_D/m_R$ and $m_R$ is the scale of new 
physics.  In this minimal scenario, new physics enters only through $m_R$, 
hence there is no other observable effect except for a nonzero $m_\nu$. 
Actually, $m_D/m_R$ is in principle observable but it is in practice far too 
small.

In general, the mass matrix of Eq.~(2) yields two nondegenerate interacting 
Majorana neutrinos (unless $m_L = m_R = 0$ is maintained exactly).  If both 
eigenvalues are small, the effective number of neutrinos counted in Big Bang 
Nucleosynthesis may be as high as six, instead of the usual three, 
depending on the mass splitting and mixing in each case~\cite{4}.

The smallness of neutrino masses may be indicative of their radiative origin. 
Many papers have been written on the subject.  For a brief review, see 
Ref.~5.  There are three one-loop mechanisms: the exchange of two scalar 
bosons with one fermion mass insertion; the exchange of one scalar boson 
with three fermion mass insertions; and the coupling to a scalar boson 
which gets a radiative vacuum expectation value through a fermion loop 
with five mass insertions.  A prime example of the first mechanism is 
the Zee model~\cite{6}.  Here the minimal standard model is extended to 
include a charged scalar singlet $\chi^+$ and a second scalar doublet 
$(\eta^+, \eta^0)$.  We then have the coupling
\begin{equation}
f_{ij} \chi^+ (\nu_i l_j - l_i \nu_j),
\end{equation}
which by itself would require $\chi^+$ to have lepton number $-2$.  However, 
this model also allows the cubic scalar coupling
\begin{equation}
\chi^- (\phi^+ \eta^0 - \phi^0 \eta^+),
\end{equation}
hence lepton number is broken explicitly.  A radiative Majorana mass matrix 
is thus obtained through the exchange and mixing of $\chi^+$ and the 
physical linear combination of $\phi^+$ and $\eta^+$.  Since $f_{ij}$ of 
Eq.~(4) is zero for $i = j$ and $\phi^+$ couples $\nu_i$ to $l_i$ with 
strength proportional to $m_{l_i}$ which is also the one fermion mass 
insertion required, the $3 \times 3$ neutrino mass matrix for $\nu_e$, 
$\nu_\mu$ and $\nu_\tau$ is of the form
\begin{equation}
{\cal M}_\nu \propto \left[ \begin{array} {c@{\quad}c@{\quad}c} 0 & 0 & 
f_{e \tau} m_\tau^2 \\ 0 & 0 & f_{\mu \tau} m_\tau^2 \\ f_{e \tau} m_\tau^2 
& f_{\mu \tau} m_\tau^2 & 0 \end{array} \right] + {\cal O} (m_\mu^2).
\end{equation}
This means that $\nu_\tau$ is almost degenerate with a linear combination 
of $\nu_\mu$ and $\nu_e$ in this model.  This may have a practical 
application in present neutrino-oscillation phenomenology~\cite{7}.

There are also three two-loop mechanisms: the exchange of three scalar bosons 
which are tied together by a cubic coupling; the exchange of two $W$ bosons; 
and the exchange of $W_L$ and $W_R$ which mix at the one-loop level.  The 
second mechanism~\cite{8} is unique in that it requires only one additional 
$\nu_R$ beyond the standard model.  In this specific case, one $\nu_L$ gets 
a seesaw mass and the other two get two-loop masses proportional to this mass 
and as functions of the charged-lepton masses with double GIM 
suppression~\cite{9}.  A detailed analytical and numerical study of this 
mechanism has been made~\cite{10}.

Finally let me return to the triplet-Higgs mechanism.  If lepton number is 
violated explicitly by the coupling of $\xi$ to the scalar doublet $\phi$, 
then one may let $\xi$ be very heavy and integrate it out to obtain the 
following effective nonrenormalizable interaction:
\begin{equation}
{1 \over M} [\phi^0 \phi^0 \nu_i \nu_j + \phi^+ \phi^0 (\nu_i l_j + l_i \nu_j) 
+ \phi^+ \phi^+ l_i l_j] + h.c.
\end{equation}
For $M \sim 10^{13}$ GeV, one gets $m_\nu \sim$ few eV.  This is the most 
economical solution and could also be a realistic model of 
leptogenesis~\cite{11} in the early universe which gets converted at the 
electroweak phase transition into the present observed baryon asymmetry.

\section{Neutrino Oscillations}

Present experimental evidence for neutrino oscillations~\cite{12} includes 
the solar $\nu_e$ deficit which requires $\Delta m^2$ of around $10^{-5}$ 
eV$^2$ for the MSW explanation or $10^{-10}$ eV$^2$ for the 
vacuum-oscillation solution, the atmospheric neutrino deficit in the 
ratio $\nu_\mu + \bar \nu_\mu / \nu_e + \bar \nu_e$ which implies a 
$\Delta m^2$ of around $10^{-2}$ eV$^2$, and the LSND experiment which 
indicates a $\Delta m^2$ of around 1 eV$^2$.  Three different $\Delta m^2$ 
necessitate four neutrinos, but the invisible width of the $Z$ boson as well 
as Big Bang Nucleosynthesis allow only three.  If all of the above-mentioned 
experiments are interpreted correctly as due to neutrino oscillations, we 
are faced with a theoretical challenge in trying to understand how 
three can equal four.  I will focus on addressing this issue rather than 
trying to review the many theoretical models for the three known neutrinos.

\section{Three Neutrinos and One Light Singlet}

One possibility is that there is a light singlet neutrino $\nu_S$ in addition 
to the three known doublet neutrinos $\nu_e$, $\nu_\mu$, and $\nu_\tau$. 
This is necessary so that it is not counted in the effective number of 
neutrinos in $Z$ decay~\cite{2}.  On the other hand, it has to mix with 
the doublet neutrinos for it to be relevant to oscillation experiments. 
Hence it is also contrained~\cite{4} by Big Bang Nucleosynthesis.  Using 
all available data, a model-independent analysis~\cite{13} shows that 
the $4 \times 4$ neutrino mass matrix must separate approximately into 
two blocks: one for $\nu_e - \nu_S$ and the other for $\nu_\mu - \nu_\tau$, 
the latter with large mixing.

An example of a specific model of this kind already exists~\cite{14}.  The 
neutrino interaction eigenstates are related to the mass eigenstates by
\begin{equation}
\left[ \begin{array} {c} \nu_S \\ \nu_e \\ \nu_\mu \\ \nu_\tau \end{array} 
\right] = \left[ \begin{array} {c@{\quad}c@{\quad}c@{\quad}c} -s & c & 
s''/\sqrt 2 & s''/\sqrt 2 \\ c & s & -s'/\sqrt 2 & s'/\sqrt 2 \\ -s' & 0 & 
-1/\sqrt 2 & 1/\sqrt 2 \\ 0 & -s'' & 1/\sqrt 2 & 1/\sqrt 2 \end{array} 
\right] \left[ \begin{array} {c} \nu_1 \\ \nu_2 \\ \nu_3 \\ \nu_4 \end{array} 
\right],
\end{equation}
where $m_1 = 0$, $m_2 \sim 2.5 \times 10^{-3}$ eV, $m_3 \sim m_4 \sim 2.5$ 
eV, with $\Delta m_{34}^2 \sim 1.8 \times 10^{-2}$ eV$^2$; $s \sim s' \sim 
0.04$, but $s''$ is undetermined.  Note that $m_{\nu_e} < m_{\nu_S}$ 
is necessary for the MSW solution~\cite{15} of the solar neutrino deficit.
Note also that $\nu_\mu$ and $\nu_\tau$ are pseudo-Dirac partners, hence 
the mixing angle for atmospheric neutrino oscillations is 45 degrees.

What is the nature of this light singlet? and how does it mix with the 
usual neutrinos?  There have been some discussions on these questions in 
the past two or three years.  One idea~\cite{16} is that it is the fermion 
partner of the massless Goldstone boson of a sponatneously broken global 
symmetry, such as lepton number (hence a Majorino) in supersymmetry.  
Another~\cite{17} is that it is the fermion partner of a scalar field 
corresponding to a flat direction (hence a modulino) in the supersymmetric 
Higgs potential.  If the standard model is extended to include a mirror 
$[SU(2) \times U(1)]'$ sector, then $\nu_S$ may be identified as a mirror 
neutrino, either in a theory where the mirror symmetry is broken~\cite{18} 
or one where it is exact~\cite{19}.  In the latter case, maximal mixing 
between the three known neutrinos and their mirror counterparts would occur 
and Big Bang Nucleosynthesis would count six neutrinos under normal conditions.

Both questions can be answered naturally in a model~\cite{20} based on 
$E_6$ inspired by superstring theory.  In the fundamental {\bf 27} 
representation of $E_6$, outside the 15 fields belonging to the minimal 
standard model, there are 2 neutral singlets.  One ($N$) is identifiable 
with the right-handed neutrino because it is a member of the {\bf 16} 
representation of $SO(10)$; the other ($S$) is a singlet also under $SO(10)$. 
In the reduction of $E_6$ to $SU(3)_C \times SU(2)_L \times U(1)_Y$, an extra 
U(1) gauge factor may survive down to the TeV energy scale.  It could be chosen 
such that $N$ is trivial under it, but $S$ is not.  This means that $N$ 
is allowed to have a large Majorana mass so that the usual seesaw mechanism 
works for the three doublet neutrinos.  At the same time, $S$ is protected 
from having a mass by the extra U(1) gauge symmetry, which I call $U(1)_N$. 
However, it does acquire a small mass from an analog of the usual seesaw 
mechanism because it can couple to doublet neutral fermions which are 
present in the {\bf 27} of $E_6$ outside the {\bf 16} of $SO(10)$.  
Renaming $S$ as $\nu_S$, the $3 \times 3$ mass matrix spanning $\nu_S$, 
$\nu_E$, and $N_E^c$ is given by
\begin{equation}
{\cal M} = \left[ \begin{array} {c@{\quad}c@{\quad}c} 0 & m_1 & m_2 \\ m_1 & 
0 & m_E \\ m_2 & m_E & 0 \end{array} \right].
\end{equation}
Hence $m_{\nu_S} \sim 2 m_1 m_2/m_E$, which is a singlet-doublet seesaw 
rather than the usual doublet-singlet seesaw.  Furthermore, the mixing of 
$\nu_S$ with the doublet neutrinos is also possible through these extra doublet 
neutral fermions.  The spontaneous breaking of $U(1)_N$ is not possible 
without also breaking the supersymmetry, hence both are assumed to occur 
at the TeV energy scale, resulting in a rich $Z'$ and Higgs 
phenomenology~\cite{21}.

\section{Three Neutrinos and One Anomalous Interaction}

If one insists on keeping only the usual three neutrinos and yet try to 
accommodate all present data, how far can one go?  It has been pointed out 
by many authors~\cite{22} that both solar and LSND data can be explained, 
as well as most of the atmospheric data except for the zenith-angle 
dependence.  It is thus worthwhile to consider the following scenario~\cite{23} 
whereby a possible anomalously large $\nu_\tau$-quark interaction may mimic 
the observed zenith-angle dependence of the atmospheric data.  Consider first 
the following approximate mass eigenstates:
\begin{eqnarray}
\nu_1 &\sim& \nu_e ~~~ {\rm with} ~ m_1 \sim 0, \\ \nu_2 &\sim& c_0 \nu_\mu 
+ s_0 \nu_\tau ~~~ {\rm with} ~ m_2 \sim 10^{-2} ~{\rm eV}, \\ \nu_3 &\sim& 
-s_0 \nu_\mu + c_0 \nu_\tau ~~~ {\rm with} ~ m_3 \sim 0.5 ~{\rm eV},
\end{eqnarray}
where $c_0 \equiv \cos \theta_0$, $s_0 \equiv \sin \theta_0$, and $\theta_0$ 
is not small.  Allow $\nu_1$ to mix with $\nu_3$ with a small angle 
$\theta '$ and the new $\nu_1$ to mix with $\nu_2$ with a small angle 
$\theta$, then the LSND data can be explained with $\Delta m^2 \sim 0.25$ 
eV$^2$ and $2 s_0 s' c' \sim 0.19$ and the solar data can be understood as 
follows.

Consider the basis $\nu_e$ and $\nu_\alpha \equiv c_0 \nu_\mu + s_0 \nu_\tau$. 
Then
\begin{equation}
-i {d \over {dt}} |\nu \rangle_{e,\alpha} = \left( p + {{\cal M}^2 \over {2p}} 
\right) |\nu \rangle_{e,\alpha},
\end{equation}
where
\begin{equation}
{\cal M}^2 = {\cal U} \left[ \begin{array} {c@{\quad}c} 0 & 0 \\ 0 & m_2^2 
\end{array} \right] {\cal U}^\dagger + \left[ \begin{array} {c@{\quad}c} 
A+B & 0 \\ 0 & B+C \end{array} \right].
\end{equation}
In the above, $A$ comes from the charged-current interaction of $\nu_e$ with 
$e$, $B$ from the neutral-cuurent interaction of $\nu_e$ and $\nu_\alpha$ with 
the quarks and electrons, and $C$ from the assumed anomalous $\nu_\tau$-quark 
interaction.  Let
\begin{equation}
{\cal U} = \left( \begin{array} {c@{\quad}c} c & s \\ -s & c \end{array} 
\right),
\end{equation}
then the resonance condition is
\begin{equation}
m_2^2 \cos 2 \theta - A + C = 0,
\end{equation}
where~\cite{24}
\begin{equation}
A - C = 2 \sqrt 2 G_F (N_e - s_0^2 \epsilon'_q N_q) p.
\end{equation}
In order to have a large $\epsilon'_q$ and yet satisfy the resonance condition 
for solar-neutrino flavor conversion, $m_2$ should be larger than its 
canonical value of $2.5 \times 10^{-3}$ eV, and $\epsilon'_q$ should be 
negative. [If $\epsilon'_q$ comes from $R$-parity violating squark exchange, 
then it must be positive, in which case an inverted mass hierarchy, {\it i.e.} 
$m_2 < m_1$ would be needed.  If it comes from vector exchange, it may be of 
either sign.]  Assuming as a crude approximation that $N_q \simeq 4 N_e$ in 
the sun, the usual MSW solution with $\Delta m^2 = 6 \times 10^{-6}$ eV$^2$ 
is reproduced here with
\begin{equation}
s_0^2 \epsilon'_q \simeq -3.92 = -4.17 (m_2^2/10^{-4}{\rm eV}^2) + 0.25.
\end{equation}
The seemingly arbitrary choice of $\Delta m_{21}^2 \sim 10^{-4}$ eV$^2$ is 
now sen as a reasonable value so that $\epsilon'_q$ can be large enough to 
be relevant for the following discussion on the atmospheric neutrino data.

Atmospheric neutrino oscillations occur between $\nu_\mu$ and $\nu_\tau$ 
in this model with $\Delta m^2_{32} \sim 0.25$ eV$^2$, the same as for 
the LSND data, but now it is large relative to the $E/L$ ratio of the 
experiment, hence the factor $\cos \Delta m^2 (L/2E)$ washes out and
\begin{equation}
P_0 (\nu_\mu \rightarrow \nu_\mu) = 1 - {1 \over 2} \sin^2 2 \theta_0 
\simeq 0.66 ~~ {\rm for} ~~ s_0 \simeq 0.47.
\end{equation}
In the standard model, this would hold for all zenith angles.  Hence it 
cannot explain the present experimental evidence that the depletion is more 
severe for neutrinos coming upward to the detector through the earth than for 
neutrinos coming downward through only the atmosphere.  This zenith-angle 
dependence appears also mostly in the multi-GeV data and not in the sub-GeV 
data.  It is this trend which determines $\Delta m^2$ to be around 
$10^{-2}$ eV$^2$ in this case.  As shown below, the hypothesis that $\nu_\tau$ 
has anomalously large interactions with quarks will mimic this zenith-angle 
dependence even though $\Delta m^2$ is chosen to be much larger, {\it i.e.} 
0.25 eV$^2$.

Consider the basis $\nu_\mu$ and $\nu_\tau$.  Then the analog of Eq.~(13) 
holds with Eq.~(14) replaced by
\begin{equation}
{\cal M}^2 = {\cal U}_0 \left[ \begin{array} {c@{\quad}c} 0 & 0 \\ 0 & m_3^2 
\end{array} \right] {\cal U}_0^\dagger + \left[ \begin{array} {c@{\quad}c} 
B & 0 \\ 0 & B + C \end{array} \right].
\end{equation}
The resonance condition is then
\begin{equation}
m_3^2 \cos 2 \theta_0 + C = 0,
\end{equation}
where $N_q$ in $C$ now refers to the quark number density inside the earth 
and the factor $s_0^2$ in Eq.~(17) is not there.  If $C$ is large enough, 
the probability $P_0$ would not be the same as the one in matter.  Using 
the estimate $N_q \sim 9 \times 10^{30}$ m$^{-3}$ and defining
\begin{equation}
X \equiv \epsilon'_q E_\nu/(10 ~{\rm GeV}),
\end{equation}
the effective mixing angles in matter are given by
\begin{eqnarray}
\tan 2 \theta_m^E &=& {{\sin 2 \theta_0} \over {\cos 2 \theta_0 + 0.091 X}} 
~~ {\rm for} ~ \nu, \\ \tan 2 \bar \theta_m^E &=& {{\sin 2 \theta_0} \over 
{\cos 2 \theta_0 - 0.091 X}} ~~ {\rm for} ~ \bar \nu.
\end{eqnarray}
For sub-GeV neutrinos, $X$ is small so matter effects are not very important. 
For multi-GeV neutrinos, $X$ may be large enough to satisfy the resonance 
condition of Eq.~(21).  Assuming adiabaticity, the neutrino and antineutrino 
survival probabilities are given by
\begin{eqnarray}
P(\nu_\mu \rightarrow \nu_\mu) &=& {1 \over 2} (1 + \cos 2 \theta_0 \cos 2 
\theta_m^E), \\ \bar P(\bar \nu_\mu \rightarrow \bar \nu_\mu) &=& {1 \over 2} 
(1 + \cos 2 \theta_0 \cos 2 \bar \theta_m^E).
\end{eqnarray}
Since $\sigma_\nu \simeq 3 \sigma_{\bar \nu}$, the observed ratio of 
$\nu + \bar \nu$ events is then
\begin{equation}
P_m \simeq {{3 r P + \bar P} \over {3 r + 1}},
\end{equation}
where $r$ is the ratio of the $\nu_\mu$ to $\bar \nu_\mu$ flux in the upper 
atmosphere.  The atmospheric data are then interpreted as follows.  For 
neutrinos coming down through only the atmosphere, $P_0 = 0.66$ applies. 
For neutrinos coming up through the earth, $P_m \simeq P_0 \simeq 
0.66$ as well for the sub-GeV data.  However, for the multi-GeV data, 
if $X = -15$, then $P = 0.31$ and $\bar P = 0.76$, hence $P_m$ is lowered 
to 0.39 if $r = 1.5$ or 0.42 if $r = 1.0$.  The apparent zenith-angle 
dependence of the data may be explained.

\section{Conclusion and Outlook}

If all present experimental indications of neutrino oscillations turn out to 
be correct, then either there is at least one sterile neutrino beyond the 
usual $\nu_e$, $\nu_\mu$, and $\nu_\tau$, or there is an anomalously large 
$\nu_\tau$-quark interaction.  The latter can be tested at the forthcoming 
Sudbury Neutrino Observatory (SNO) which has the capability of neutral-current 
detection.  The predicted $\Delta m^2$ of 0.25 eV$^2$ in $\nu_\mu$ to $\nu_e$ 
and $\nu_\tau$ oscillations will also be tested at the long-baseline neutrino 
experiments such as Fermilab to Soudan 2 (MINOS), KEK to Super-Kamiokande 
(K2K), and CERN to Gran Sasso.

More immediately, the new data from Super-Kamiokande, Soudan 2, and MACRO on 
$\nu_\mu + \bar \nu_\mu$ events through the earth should be analyzed for 
such an effect.  For a zenith angle near zero, the $\Delta m^2 \sim 10^{-2}$ 
eV$^2$ oscillation scenario should have $R \sim 1$, whereas the $\Delta m^2 
\sim 0.25$ eV$^2$ oscillation scenario (with anomalous interaction) would 
have $R = P_0 \sim 0.66$.  Furthermore, if $\nu$ and $\bar \nu$ can be 
distinguished (as proposed in the HANUL experiment), then to the extent 
that $CP$ is conserved, matter effects can be isolated.

Neutrino physics is on the verge of major breakthroughs.  New experiments 
in the next several years will be decisive in leading us forward in its 
theoretical understanding, and may even discover radically new physics 
beyond the standard model.

\section*{Acknowledgments} I thank Biswarup Mukhopadhyaya and all the other 
organizers for their great hospitality and a stimulating workshop.  This work 
was supported in part by the U.~S.~Department of Energy under Grant 
No.~DE-FG03-94ER40837.

\end{document}